\documentclass[aps,floatfix,10pt,showpacs,prd,superscriptaddress,twocolumn]{revtex4-2}

\usepackage{graphicx}	
\usepackage{amsmath}	
\usepackage{amssymb}	
\usepackage{xcolor} 

\newcommand{\bham}{
\affiliation{Institute for Gravitational Wave Astronomy 
\& School of Physics and Astronomy, University of 
Birmingham, Birmingham, B15 2TT, UK}}
\newcommand{\ral}{
\affiliation{Science and Technology Facilities Council,
Rutherford Appleton Laboratory, 
Harwell Campus, Didcot, OX11 0QX, UK}}
\newcommand{\Livingston}{
\affiliation{LIGO Livingston Observatory, Livingston, LA 70754, United States of America}}

\begin{document}

\title{Sensors and Actuators for the Advanced LIGO+ Upgrade}


\author{S J Cooper}
\bham

\author{C M Mow-Lowry}
\email{Now at the Vrije Universiteit Amsterdam}
\bham

\author{D Hoyland}
\bham

\author{J Bryant}
\bham
\author{A Ubhi}
\bham

\author{J O’Dell}
\ral

\author{A Huddart}
\ral 

\author{S Aston}
\Livingston

\author{A Vecchio}
\email{av@star.sr.bham.ac.uk}
\bham

\date{\today}




\begin{abstract}
As part of the Advanced LIGO+ (A+) project we have developed, produced, and characterised sensors and electronics to interrogate new optical suspensions. The central element is a displacement sensor with an integrated electromagnetic actuator known as a BOSEM and its readout and drive electronics required to integrate them into LIGO's control and data system. In this paper we report on improvements to the sensors and testing procedures undertaken to meet enhanced performance requirements set out by the A+ upgrade to the detectors. The best devices reach a noise level of 4.5$\times 10^{-11}$\,m/$\sqrt{\rm Hz}$ at a measurement frequency of 1\,Hz. 

\vspace{0.1cm}

\end{abstract}

\maketitle 

\section{Introduction}

The Advanced LIGO~\cite{LIGO03} and Virgo~\cite{acernese2021calibration} gravitational-wave interferometers have opened explorations of the gravitational-wave sky, with the observation of the coalescence of stellar-mass binary black holes~\cite{gw150914, O1-BBH, gwtc-1, gwtc-2, gwtc-2.1, gwtc-3},  binary neutron stars~\cite{gw170817, gw170817-MM, gw190425} and more recently neutron star-black hole systems~\cite{NS-BH-O3}. 



The LIGO and Virgo detectors are being upgraded to what is known as `Advanced$+$' (A$+$, hereafter) configuration. Installation has already begun for some of the hardware. After the forth observing run (O4) scheduled to start in 2022, it will be completed in advance of the fifth observing run in 2025~\cite{ObservingScenario}. The higher sensitivity achieved through this upgrade will correspond to an increase of the Universe's observable volume by a factor $\approx 5$ with respect to the one probed by the existing instruments, resulting in an equal increase in detection rate. For example, based on current estimates of the merger rate of populations of stellar-mass binary systems~\cite{gwtc-3-rates-pop} and the expected instrument performance in A$+$ configurations, the detectors will observe several of such binary coalescences every week, \textit{e.g.}~\cite{PhysRevD.91.062005}.

LIGO A$+$ will achieve this increase in observing range by upgrading several sub-systems of the detectors: new test-masses with reduced coating thermal noise \cite{abernathy2021exploration}, frequency-dependent squeezing \cite{PhysRevLett.124.171102}, replacement the central beam-splitter with a larger optic, and installation of balanced homodyne detection \cite{Fritschel:14} to further reduce quantum noise. 

\begin{figure}[!h]
  \includegraphics[width=0.3\textwidth]{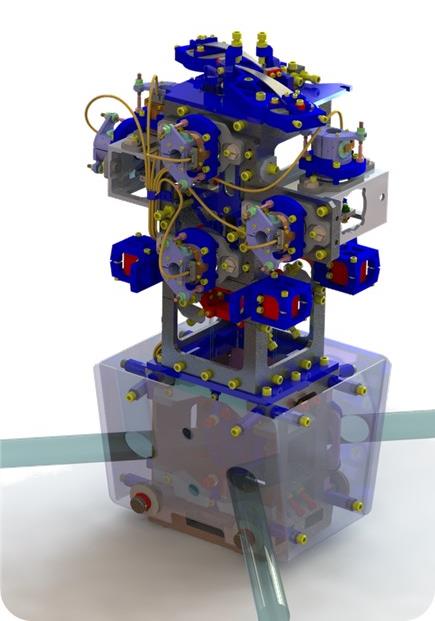}
  \caption{A rendered image of the new HAM Relay Tiple Suspension with BOSEMs installed.}
  \label{fig:hrts}
\end{figure}

These upgrades will require several new multi-stage suspension systems for isolating and controlling the new optical elements. Each suspension requires a set of low-noise sensors and actuators, with associated electronics, to actively damp resonances and steer the laser beam. One example of a new suspension, shown in Fig.~\ref{fig:hrts}, is the `HAM Relay Triple Suspension', an evolution of existing suspension designs that is lighter and easier to assemble. In total, 200 BOSEMs -- "Birmingham Optical Sensor and Electro-Magnetic actuator" -- have been produced, along with 82 coil-driver units and 44 `satellite amplifiers' for reading out the photocurrent.

This paper updates the design produced for Advanced LIGO \cite{Carbone_2012, Aston_2012} to fulfil existing and enhanced sensitivity requirements for A+. A noise budget with the dominant terms is provided, with reference to the exhaustive selection procedure required. More than half of the BOSEMs reach a new `Enhanced' performance requirement. The best Enhanced BOSEMs are now dominated by quantum shot noise across the whole measurement band, a major performance improvement.

\section{BOSEM design}
BOSEMs are compact, ultra-high vacuum compliant, non-contact, and low-noise position sensors with integrated electromagnetic actuators. They have a long history of development started by studies on the initial LIGO OSEMS by Fritschel \cite{fritschelOSEM}, which were upgraded to AOSEMS by Abbott \cite{abbott2009advanced} and modified by Carbone \cite{Carbone_2012} to produce the Advanced LIGO BOSEMs. A table of key parameters for each BOSEM can be found in Table \ref{tab:bosemParam}.

\begin{figure}[!h]
  \includegraphics[width=0.5\textwidth]{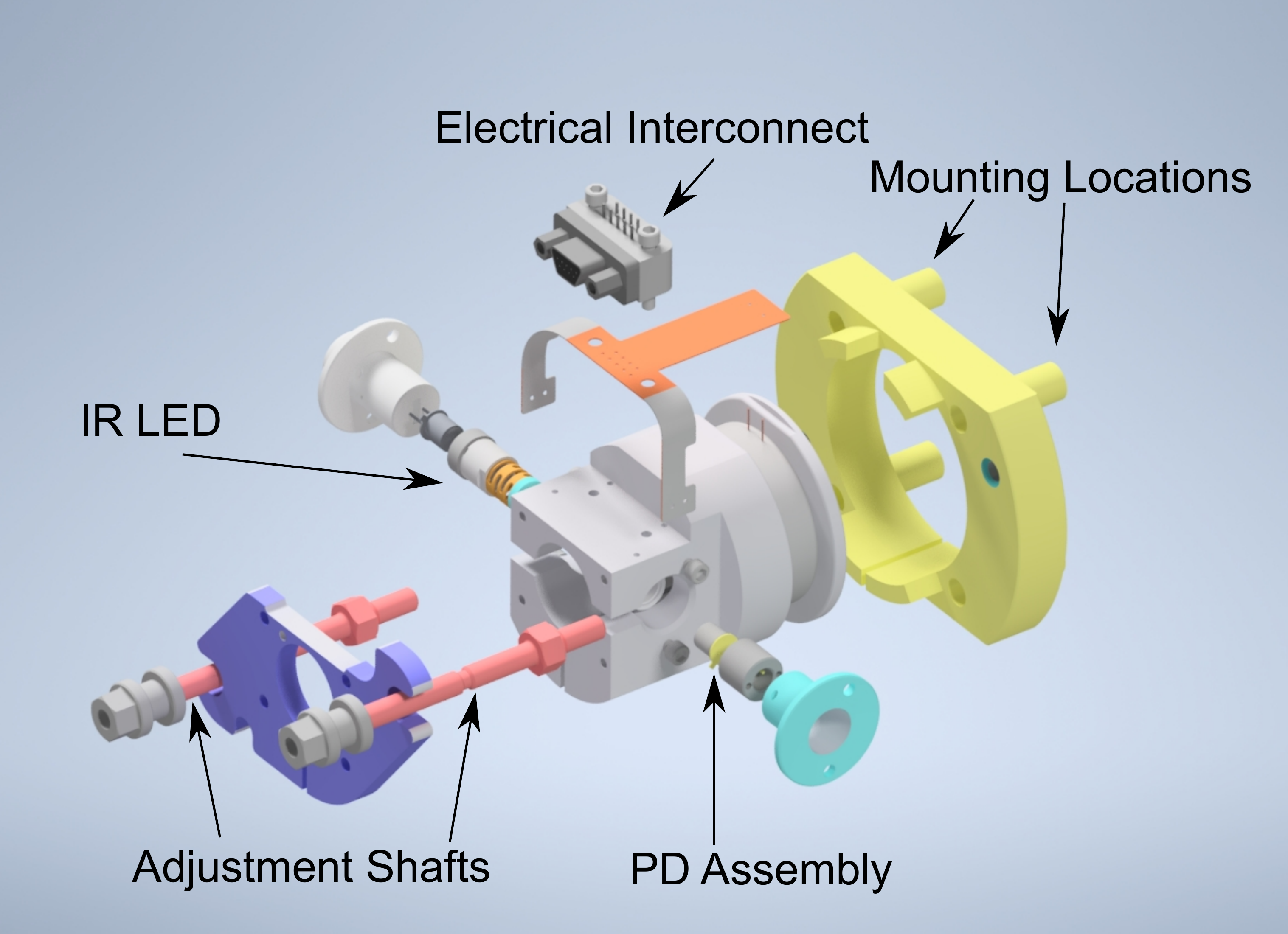}
  \caption{An exploded view of a BOSEM highlighting key components.}
  \label{fig:explodedBosem}
\end{figure}

Each BOSEM unit comprises of a number of elements: sensing, actuation, and alignment. An exploded CAD model of the BOSEM is shown in Fig.\,\ref{fig:explodedBosem}, with key features of the sensor labelled. The optical readout is based on a shadow sensing scheme where an opaque `flag', rigidly mounted to the measurement surface \cite{BOSEMflag}, partly blocks 935\,nm light emitted from an Infra-Red LED (IRLED, Optek OP132) before it is sensed by a photodiode (PD, Osram BPX65). The choice of wavelength ensures high quantum efficiency from the silicon photodiode and negligible emission at 1064\,m, the wavelength of the main science laser. A lens is installed after the LED to produce a narrower beam and smaller spot on the photodiode, improving both the linearity and sensitivity of the BOSEM. 

\begin{table}[]
  \caption{Key parameters for a BOSEM.}
  \label{tab:bosemParam}
  \begin{tabular}{|c|c|}
  \hline
  Parameter & Value \\ \hline
  \multicolumn{2}{|c|}{Coil} \\ \hline
  Turns & 800 \\
  Winding sense & \begin{tabular}[c]{@{}c@{}}Clockwise when viewed \\ from the rear face\end{tabular} \\
  Inductance & $14.7\pm 0.7$\,mH \\
  Resistance & $37.6 \pm 2\,\Omega$ \\
  Length & 8\,mm \\
  Inner coil diameter & 17.8\,mm \\
  Maximum current & 150\,mA \\
  Breakdown (to coil former) & \textgreater 200\,V \\ \hline
  \multicolumn{2}{|c|}{Sensor} \\ \hline
  Mass & 158\,g \\
  Linear range (typ.) & 0.7\,mm \\
  Operating LED current & 35\,mA \\
  Photodiode bias & 10\,V typ. (50\,V max.) \\
  \begin{tabular}[c]{@{}c@{}}Photocurrent, open-light \end{tabular} & 45-80\,$\mu$A \\
  Current transfer ratio & $0.19 \pm 0.04\%$ \\
  Average sensitivity & 20.25\,kV/m \\
  Electrical connector & \begin{tabular}[c]{@{}c@{}}Glenair micro-D \\  MR7590-9P-1BSN-MC225\end{tabular} \\
  Mechanical Interface & \begin{tabular}[c]{@{}c@{}} 4x 8/32UNC thru-holes on \\ 40.64\,mm (1.3’’) square grid\end{tabular} \\
  Operating temperature & $22\pm2^\circ$\,C \\
  \begin{tabular}[c]{@{}c@{}}Storage temperature\\ (ambient pressure)\end{tabular} & 10 to 120°C \\ \hline
  \multicolumn{2}{|c|}{Sensor Noise} \\ \hline
  \begin{tabular}[c]{@{}c@{}} Standard \\ 1-10\,Hz \\ 10-20\,Hz\end{tabular} & \begin{tabular}[c]{@{}c@{}} \\ 3x10$^{-10}$\,m/$\sqrt{\rm Hz}$\\ 1x10$^{-10}$\,m/$\sqrt{\rm Hz}$\end{tabular} \\ \hline
  \begin{tabular}[c]{@{}c@{}} Enhanced \\ 1-20\,Hz\end{tabular} &  
  \begin{tabular}[c]{@{}c@{}} \\ \textless\,0.75x10$^{-10}$\,m/$\sqrt{\rm Hz}$ \end{tabular}\\ \hline
  \end{tabular}
  \end{table}

To meet LIGO's stringent vacuum requirements, the BOSEMs must go through a multi-stage cleaning process. In the original production run toluene was used in the cleaning process to remove paraffin from the coil wire \cite{wireClean}. Following the detection of toluene residue in RGA scans during the production process the coil wire was switched(from MWS Wire 32 HML to MWS Wire 32 HML Natural) to ensure the production process was paraffin-free. The updated cleaning process can be found in \citep{BOSEMClean}.

\section{Performance analysis}

\begin{figure}[]
  \centering
  \includegraphics[width=1\linewidth]{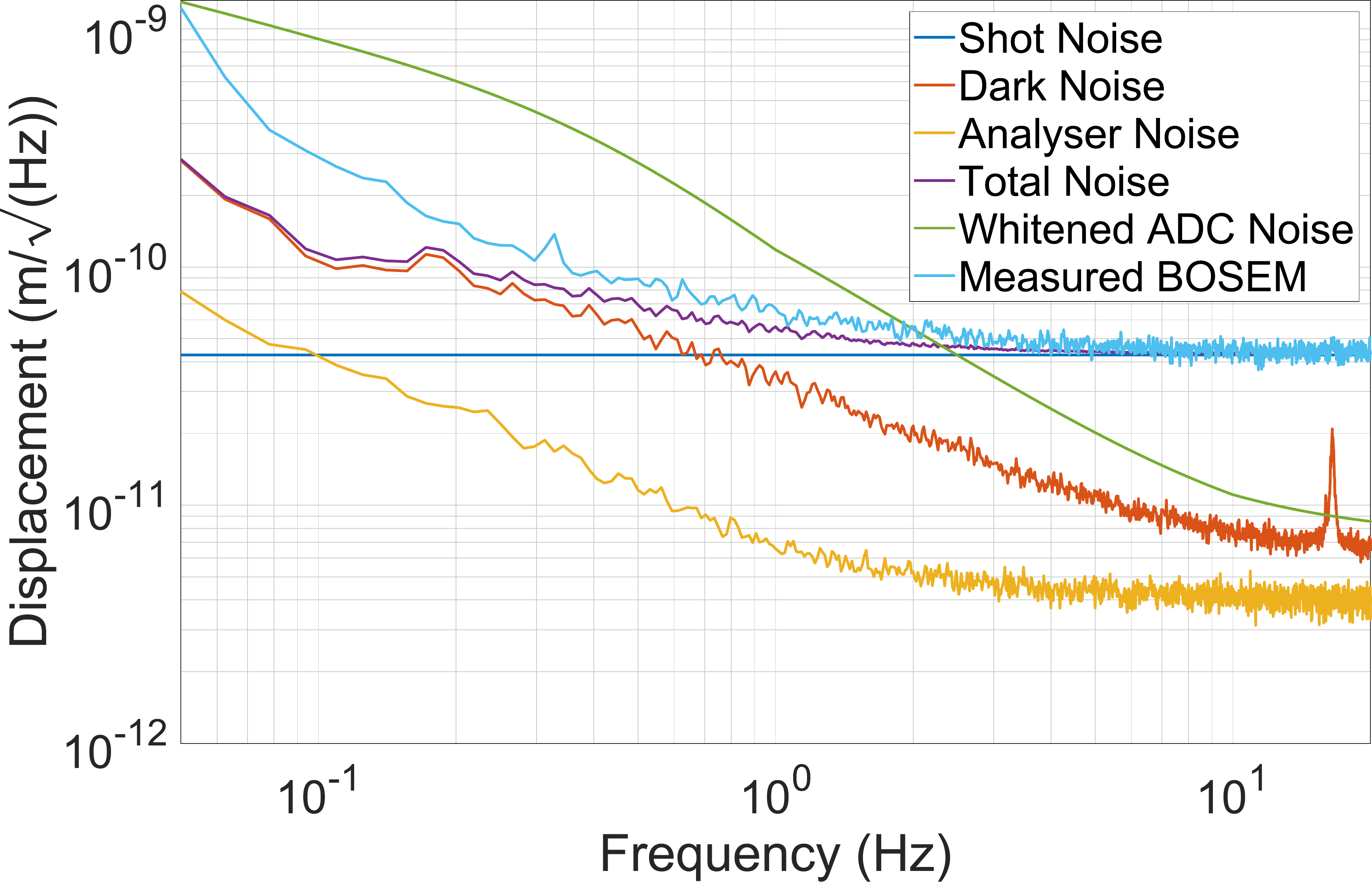}
  \caption{A noise budget of an Enhanced BOSEM showing each of the individual noise contributions. The main noise sources are the shot noise (dark blue) and the photodiode dark noise (red). The resolution of the BOSEM at the critical 10\,Hz measurement frequency is limited by optical shot noise.}
  \label{fig:bosemNB}
\end{figure}

The noise budget of a BOSEM is shown in Fig.~\ref{fig:bosemNB}. The noise sources expected to dominate a typical system are the shot noise and the photodiode dark noise. The shot noise, shown in dark blue, is modelled and projected to an equivalent displacement using
\begin{equation}
SN = \frac{2R}{K} \sqrt{2eI} 
\end{equation}
where $e$ is the electronic charge, $I$ is the photocurrent at the operating point (the `half-light' current), $R$ is the transimpedance gain of 121\,kΩ, and $K$ is the sensivity of the BOSEM which varies depending on the photocurrent but averages 20250 volts per metre. The leading factor of 2 accounts for the differential output gain of the satellite amplifier. The photodiode dark noise, shown in red, was measured by switching off the IRLED and measuring the output of the satellite amplifier \cite{Carbone_2012}. 

The total modelled noise trace, shown in purple, is achieved by summing the shot noise and dark noise in quadrature. The noise model is compared with the spectrum of a typical BOSEM `open light test', where there is no flag inserted in the beam, shown in light blue. Since the open light test has a factor of 2 more photocurrent than normal half-light operation, to calibrate it into metres we divide by a factor of $\sqrt{2}$ before using the volts-to-metres conversion factor. This will correctly scale the shot-noise for the photocurrent expected under normal operating conditions.

The noise floor of our signal analyser is shown for reference in yellow. The whitened ADC noise of LIGO's Control and Data System is shown in green; at frequencies lower than 1\,Hz we expect BOSEMs installed at LIGO to be dominated by this noise source.

There is a clear discrepancy between the noise model and the measurement of a typical BOSEM. Two contributions have been identified. First, the photodiode dark noise varies from unit to unit, presumably due to defects in individual photodiodes. Repeated measurements have shown that this never exceeds the noise requirements and to be sub-dominant in almost all BOSEM assemblies. Second, the IRLEDs show significant excess intensity fluctuation, and this is the dominant source of excess noise. This excess noise was identified in the original Advanced LIGO BOSEM production run and a screening process was implemented to select IRLEDs that comply with the noise requirements \cite{AstonPhD}. For the A+ upgrade we conducted an extensive review of different IRLEDs, but no model was shown to consistently meet the noise requirements. Instead, the Advanced LIGO screening process was improved, most importantly by lengthening the initial `burn in' from 50 hours to 168 hours as detailed in \cite{BOSEMscreen}. 

\begin{figure}[]
  \centering
  \includegraphics[width=1\linewidth]{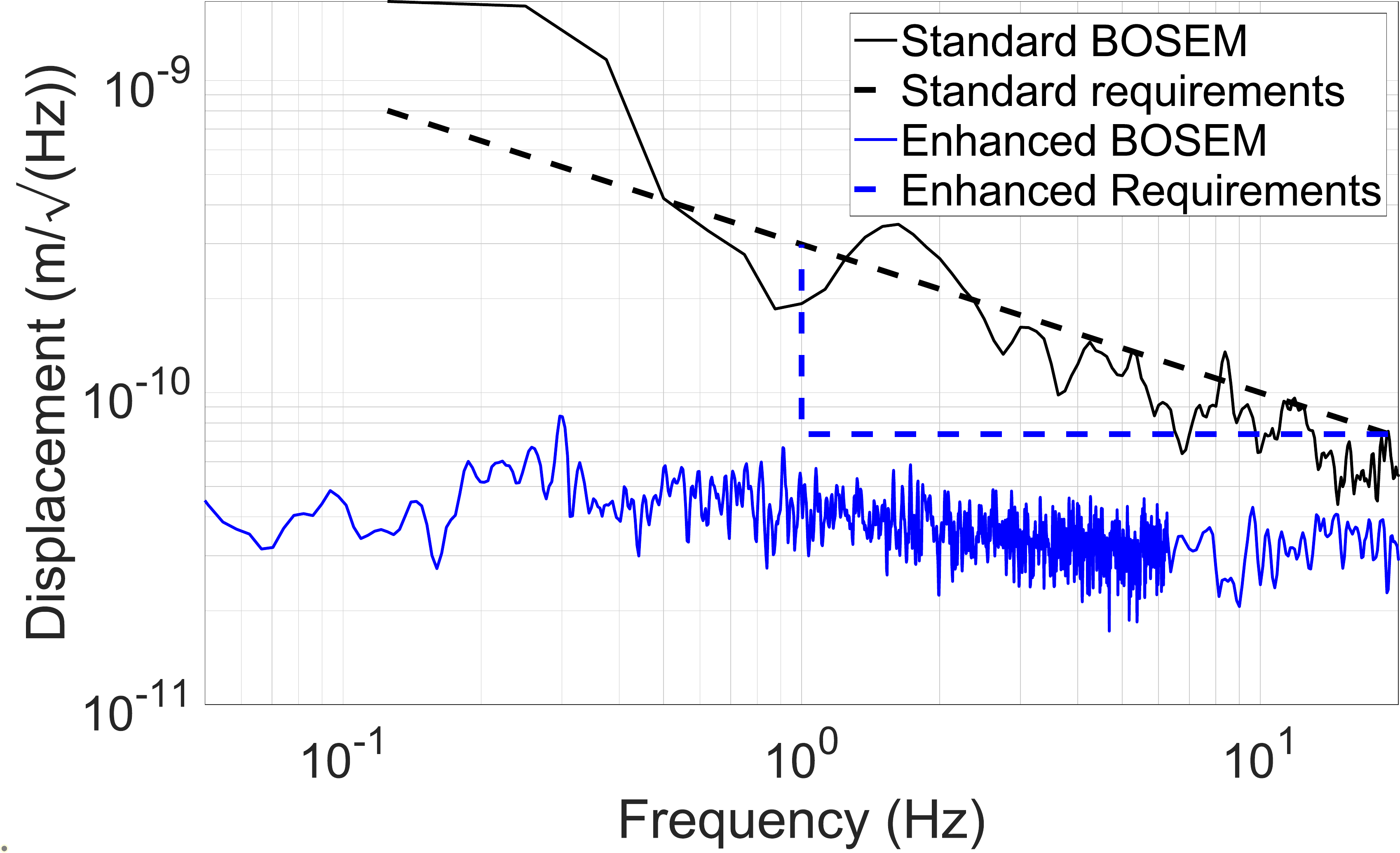}
  \caption{Amplitude Spectral density of the Standard (black) and Enhanced (blue) BOSEMs compared against the noise requirements shown with dashed lines.}
  \label{fig:bosemASD}
\end{figure}

To capitalise on the effort involved in screening every IRLED, we identified units with especially good noise performance between 1\,Hz and 20\,Hz. If these units met a new `Enhanced' noise requirement they were separated and tagged for use in critical locations. They do not require any change to the existing signal chain. Figure \ref{fig:bosemASD} shows the noise spectra of a Standard BOSEM and an Enhanced BOSEM, along with their respective requirements. The best of the Enhanced BOSEMs are dominated by shot noise across the whole measurement band. Further improvements can come from reducing the measurement range (increasing sensitivity), for example the Differential-OSEM sensors \cite{DOSEM}, or by using alternative technology such as fringe-counting interferometers \cite{HoQI}, which have improved resolution without sacrificing operating range.

\section{Conclusions}

As part of the Advanced LIGO+ upgrade the University of Birmingham has provided over 200 BOSEM sensors and actuators along with their associated driving electronics. Over half of the BOSEMs meet the `Enhanced' specification. The best units have a resolution of 4.5$\times 10^{-11}\,$m$\sqrt{\rm Hz}$ all the way down to 0.1\,Hz and these devices represent the ultimate performance of BOSEMs in their current form.

\acknowledgments
We thank Jeff Kissel for his useful comments on the manuscript. This work is supported by the UK's Science and Technology Facilities Council through Grant No. ST/S00243X/1. A.V. acknowledges the support of the Royal Society and Wolfson Foundation. The authors gratefully acknowledge the support of the United States NSF for the construction and operation of the LIGO Laboratory and Advanced LIGO. LIGO was constructed by the California Institute of Technology and Massachusetts Institute of Technology with funding from the United States National Science Foundation (NSF), and operates under cooperative agreement PHY–1764464. Advanced LIGO was built under
award PHY–0823459.

\bibliographystyle{unsrt}
\bibliography{refs}

\begin{thebibliography}{10}

\bibitem{LIGO03}
A.~Buikema et~al.
\newblock Sensitivity and performance of the {Advanced LIGO} detectors in the
  third observing run.
\newblock {\em Phys. Rev. D}, 102:062003, 2020.

\bibitem{acernese2021calibration}
F.~Acernese, M.~Agathos, A.~Ain, S.~Albanesi, A.~Allocca, A.~Amato, T.~Andrade,
  N.~Andres, T.~Andri{\'c}, S.~Ansoldi, et~al.
\newblock Calibration of advanced virgo and reconstruction of detector strain
  h(t) during the observing run {O3}.
\newblock {\em arXiv preprint arXiv:2107.03294}, 2021.

\bibitem{gw150914}
B.P. Abbott et~al.
\newblock {Observation of Gravitational Waves from a Binary Black Hole Merger}.
\newblock {\em Phys. Rev. Lett.}, 116(6):061102, 2016.

\bibitem{O1-BBH}
B.P. Abbott et~al.
\newblock {Binary Black Hole Mergers in the first Advanced LIGO Observing Run}.
\newblock {\em Phys. Rev. X}, 6(4):041015, 2016.
\newblock [Erratum: Phys.Rev.X 8, 039903 (2018)].

\bibitem{gwtc-1}
B.P. Abbott et~al.
\newblock {GWTC-1: A Gravitational-Wave Transient Catalog of Compact Binary
  Mergers Observed by LIGO and Virgo during the First and Second Observing
  Runs}.
\newblock {\em Phys. Rev. X}, 9(3):031040, 2019.

\bibitem{gwtc-2}
R.~Abbott et~al.
\newblock {GWTC-2: Compact Binary Coalescences Observed by LIGO and Virgo
  During the First Half of the Third Observing Run}.
\newblock {\em Phys. Rev. X}, 11:021053, 2021.

\bibitem{gwtc-2.1}
R.~Abbott et~al.
\newblock {GWTC-2.1: Deep Extended Catalog of Compact Binary Coalescences
  Observed by LIGO and Virgo During the First Half of the Third Observing Run}.
\newblock 8 2021.

\bibitem{gwtc-3}
R.~Abbott et~al.
\newblock {GWTC-3: Compact Binary Coalescences Observed by LIGO and Virgo
  During the Second Part of the Third Observing Run}.
\newblock 11 2021.

\bibitem{gw170817}
B.P. Abbott et~al.
\newblock {GW170817: Observation of Gravitational Waves from a Binary Neutron
  Star Inspiral}.
\newblock {\em Phys. Rev. Lett.}, 119(16):161101, 2017.

\bibitem{gw170817-MM}
B.P. Abbott et~al.
\newblock {Multi-messenger Observations of a Binary Neutron Star Merger}.
\newblock {\em Astrophys. J. Lett.}, 848(2):L12, 2017.

\bibitem{gw190425}
B.P. Abbott et~al.
\newblock {GW190425: Observation of a Compact Binary Coalescence with Total
  Mass $\sim 3.4 M_{\odot}$}.
\newblock {\em Astrophys. J. Lett.}, 892:L3, 2020.

\bibitem{NS-BH-O3}
R.~Abbott et~al.
\newblock {Observation of gravitational waves from two neutron star-black hole
  coalescences}.
\newblock {\em Astrophys. J. Lett.}, 915:L5, 2021.

\bibitem{ObservingScenario}
B.P. Abbott et~al.
\newblock {Prospects for Observing and Localizing Gravitational-Wave Transients
  with Advanced LIGO, Advanced Virgo and KAGRA}.
\newblock {\em Living Rev. Rel.}, 21(1):3, 2018.

\bibitem{gwtc-3-rates-pop}
R.~Abbott et~al.
\newblock {The population of merging compact binaries inferred using
  gravitational waves through GWTC-3}.
\newblock 11 2021.

\bibitem{PhysRevD.91.062005}
J.~Miller, L.~Barsotti, S.~Vitale, P.~Fritschel, M.~Evans, and D.~Sigg.
\newblock Prospects for doubling the range of {Advanced LIGO}.
\newblock {\em Phys. Rev. D}, 91:062005, 2015.

\bibitem{abernathy2021exploration}
M.~Abernathy, A.~Amato, A.~Ananyeva, S.~Angelova, B.~Baloukas, R.~Bassiri,
  G.~Billingsley, R.~Birney, G.~Cagnoli, M.~Canepa, et~al.
\newblock Exploration of co-sputtered {Ta$_2 $O$_5$-ZrO$_2$} thin films for
  gravitational-wave detectors.
\newblock {\em arXiv preprint arXiv:2103.14140}, 2021.

\bibitem{PhysRevLett.124.171102}
L.~McCuller, C.~Whittle, D.~Ganapathy, K.~Komori, M.~Tse, A.~Fernandez-Galiana,
  L.~Barsotti, P.~Fritschel, M.~MacInnis, F.~Matichard, K.~Mason, N.~Mavalvala,
  R.~Mittleman, Haocun Yu, M.~E. Zucker, and M.~Evans.
\newblock Frequency-dependent squeezing for {Advanced LIGO}.
\newblock {\em Phys. Rev. Lett.}, 124:171102, 2020.

\bibitem{Fritschel:14}
P.~Fritschel, M.~Evans, and V.~Frolov.
\newblock Balanced homodyne readout for quantum limited gravitational wave
  detectors.
\newblock {\em Opt. Express}, 22:4224--4234, 2014.

\bibitem{Carbone_2012}
L.~Carbone, S.~M. Aston, R.~M. Cutler, A.~Freise, J.~Greenhalgh, J.~Heefner,
  D.~Hoyland, N.~A. Lockerbie, D.~Lodhia, N.~A. Robertson, C.~C. Speake, K.~A.
  Strain, and A.~Vecchio.
\newblock Sensors and actuators for the advanced {LIGO} mirror suspensions.
\newblock {\em Classical and Quantum Gravity}, 29:115005, 2012.

\bibitem{Aston_2012}
S~M Aston, M~A Barton, A~S Bell, N~Beveridge, B~Bland, A~J Brummitt, G~Cagnoli,
  C~A Cantley, L~Carbone, A~V Cumming, L~Cunningham, R~M Cutler, R~J~S
  Greenhalgh, G~D Hammond, K~Haughian, T~M Hayler, A~Heptonstall, J~Heefner,
  D~Hoyland, J~Hough, R~Jones, J~S Kissel, R~Kumar, N~A Lockerbie, D~Lodhia,
  I~W Martin, P~G Murray, J~O'Dell, M~V Plissi, S~Reid, J~Romie, N~A Robertson,
  S~Rowan, B~Shapiro, C~C Speake, K~A Strain, K~V Tokmakov, C~Torrie, A~A van
  Veggel, A~Vecchio, and I~Wilmut.
\newblock Update on quadruple suspension design for advanced {LIGO}.
\newblock {\em Classical and Quantum Gravity}, 29:235004, 2012.

\bibitem{fritschelOSEM}
P.~Fritschel and R.~Adhikari.
\newblock Characterization and comparison of a potential new (sus) local
  sensor.
\newblock {\em {LIGO} Document Control Center}, 2015.
\newblock Available at \url{https://dcc.ligo.org/LIGO-T990089/public}.

\bibitem{abbott2009advanced}
R.~Abbott, M.~Barton, B.~Bland, B.~Moore, C.~Osthelder, and J.~Romie.
\newblock Advanced {LIGO OSEM} final design document.
\newblock {\em {LIGO} Document Control Center}, 2014.
\newblock Available at \url{https://dcc.ligo.org/LIGO-T0900286/public}.

\bibitem{BOSEMflag}
M.~Evans and M.~Hillard.
\newblock {BOSEM} flat magnet flag, {aLIGO SUS}.
\newblock {\em {LIGO} Document Control Center}, 2011.
\newblock Available at \url{https://dcc.ligo.org/LIGO-D1100573/public}.

\bibitem{wireClean}
C.~Taylor.
\newblock Cleaning procedure for magnet wire with {ML/HML} insulation.
\newblock {\em {LIGO} Document Control Center}, 2019.
\newblock Available at \url{https://dcc.ligo.org/T040127}.

\bibitem{BOSEMClean}
J.L.~Bryant S.~M.~Aston, D.~Lodhia and D.~Hoyland.
\newblock {BOSEM} assembly specification.
\newblock {\em {LIGO} Document Control Center}, 2021.
\newblock Available at \url{https://dcc.ligo.org/LIGO-T060233/public}.

\bibitem{AstonPhD}
S.~M. Aston.
\newblock {\em Optical read-out techniques for the control of test-masses in
  gravitational wave observatories}.
\newblock University of Birmingham, 2011.
\newblock Available at \url{https://etheses.bham.ac.uk/id/eprint/1665/}.

\bibitem{BOSEMscreen}
D.~Hoyland and J.~L. Bryant.
\newblock {A+ BOSEM IR-LED} screening.
\newblock {\em {LIGO} Document Control Center}, 2019.
\newblock Available at \url{https://dcc.ligo.org/LIGO-T1900596/public}.

\bibitem{DOSEM}
J.~Conklin, D.~Jariwala, T~Pechsiri, H.~Inchauspe, P.~Fulda, and D.~Tanner.
\newblock Progress in developing a {Differential OSEM (DOSEM)}.
\newblock {\em {LIGO} Document Control Center}, 2019.
\newblock Available at \url{https://dcc.ligo.org/LIGO-G1900464}, manuscript in
  preparation.

\bibitem{HoQI}
S.~J. Cooper, C.~J. Collins, A.~C. Green, D.~Hoyland, C.~C. Speake, A.~Freise,
  and C.~M. Mow-Lowry.
\newblock A compact, large-range interferometer for precision measurement and
  inertial sensing.
\newblock {\em Classical and Quantum Gravity}, 35:095007, 2018.

\end{thebibliography}

\end{document}